\documentstyle[11pt,psfig]{article}
\title{Giant resonances in exotic spherical nuclei \\
within the RPA approach with the Gogny force}

\date{}

\author{S. P\'eru${^1}{^*}$, J.F. Berger$^1$, and P.F. Bortignon$^2$\\
\small{$^1$DPTA/Service de Physique Nucl\'eaire,
 BP12 --F --91680 Bruy\`eres--le--Ch\^atel},\\
\small{$^2$Dipartimento di Fisica, Universit\`a di Milano and INFN,
Via Celoria 16, 20133 Milano, Italy}\\
$^*$corresponding author.  \\
 {\it e-mail address :} sophie.peru-desenfants@cea.fr.\\}

\normalsize
\begin{document}

\baselineskip 11pt
\def\dspt{\displaystyle}

\maketitle
\begin{abstract}
Theoretical results for giant resonances in the three doubly magic
exotic nuclei $^{78}$Ni,  $^{100}$Sn and $^{132}$Sn are obtained from
Hartree-Fock (HF) plus Random Phase Approximation (RPA) calculations
using the D1S parametrization of the Gogny two-body effective
interaction.
Special attention is paid to full consistency between the HF field and
the RPA particle-hole residual interaction.
The results for the exotic nuclei, on average, appear similar to
those of stable ones, especially for quadrupole and octupole states.
More exotic systems have to be studied in order to confirm such a
trend. The low energy of the monopole resonance in $^{78}$Ni suggests
that the compression modulus in this neutron rich nucleus is lower than
the one of stable ones.
\end{abstract}

\section{Introduction}
Giant multipole resonances (GR) are collective excitations of nuclei that
lie at excitation energies above the nucleon separation energy
(8-10 MeV), have different multipolarities and carry different
spin-isospin quantum numbers.
They have been observed for stable nuclei throughout the mass table
with large cross sections, close to the maximum allowed by sum rule
arguments, implying that a large number of nucleons participate in a
very collective nuclear motion~\cite{book1,book2}.
It is a challenge both to experimentalists and theorists to study the
properties of these states for nuclei far from the valley of
stability.
Not too much has been done from the experimental side yet:
let us just mention the two measurements of the electric dipole GR
(GDR) made in neutron-rich oxygen isotopes~\cite{GSI,MSU}.
Beside GR, there are also low-lying collective excitations, in
particular quadrupole and octupole states, which reflect much more than
the GR the detail of shell structure. More experimental data are
available for such states \cite{q2} in the case of unstable nuclei,
giving us information on the modifications of the shell structure far
from stability.

From the theoretical side,  more and more calculations of GR and
low-lying states are performed nowadays in the framework of microscopic
HF+RPA or HFB+QRPA approaches. The effective nucleon-nucleon
interactions used are taken as non-relativistic effective
two-body potentials~\cite{hiro1} or relativistic Lagrangians for meson
exchange~\cite{Vre}.
Such microscopic approaches, although less accurate than more
phenomenological ones, usually describe reasonably well the properties
of these states in stable nuclei.

Among the effective forces used in the non-relativistic approaches,
 the Gogny force~\cite{ref4,ref5} is one of those which has been
extensively employed for the description of GR and low-lying states in
doubly closed shell nuclei with the RPA method~\cite{ref2,ref6,ref3}.
Recently, this force has been used for the first time in full
Quasi-Particle RPA (QRPA) calculations. Chains of isotopes in the
oxygen, nickel and tin regions have been studied in order to derive the
properties of low-lying states~\cite{milan}.

The purpose of this paper is to present the results of calculations
performed in three spherical exotic nuclei: $^{78}$Ni,  $^{100}$Sn and
$^{132}$Sn, and to compare them with those obtained in stable nuclei.
More precisely, GR and low-lying states in these nuclei will be
analyzed and comparisons will be made with systematics and with
analogous quantities in the well-known $^{208}$Pb. The latter nucleus
will serve as a reference and, for this reason, results for $^{208}$Pb
will be displayed along with those of the three exotic nuclei in most
Tables and Figures.
Let us point out that the results presented here for $^{208}$Pb are new.
They have been derived with the D1S parameterization of the Gogny force
which is the one currently used now. They slightly differ from those of
Ref.~\cite{ref2} where the older parameterization D1 was employed.

A point we pay special attention to in the present work  is the
effect of the full consistency of the residual particle-hole (p-h)
interaction with the mean field produced by the same force, as allowed
by the use of consistently combined HF and RPA approaches.
 In order to analyze this effect, we present results where
different components of the residual p-h interaction such as those
generated by the spin-orbit or the Coulomb force are switched off. As
will be seen, the influence of these often omitted components are far
from being negligible.

 In the following Section details concerning the parameters of the
two-body force, the numerical methods used for solving the RPA
equations are briefly recalled along with a few useful formulas.
Results are presented and discussed in Section~3. The main conclusions
of this work are summarized in Section~4.
Let us mention that a preliminary account of the present results
has appeared in the workshop Proceedings of Ref.~\cite{curie}.

\section{The HF+RPA approach with the Gogny force}

The RPA approach employed here is described in Refs.~\cite{ref6,ref2,ref3}.
The effective force D1S proposed by Gogny~\cite{ref4,ref5} is used.
This finite-range density-dependent interaction  describes the mean field
of the nucleus, and the residual interaction in the RPA calculations is
obtained via the functional  second derivative of the mean field
with respect to the one-body density matrix.
We want to stress that all the terms of the effective force are
considered in the HF mean-field and in the residual p-h interaction,
including  the spin--spin component, the Coulomb force and the
terms produced by the two-body spin-orbit interaction.
Only the two-body terms coming from the two-body center of mass
correction are not included in the RPA matrix elements.
Therefore, they have been also left out from the mean field
calculations.
In order to get equivalent binding  energies and radii, the
coefficient of the spin-orbit component of D1S has been reduced from
130 MeV to 115 MeV.  Such a procedure was previously employed in
calculations with the D1 force, as explained in Ref.~\cite{ref4}.
The Gogny force D1S including this change of the spin-orbit strength
will be called D1S'.

 In the results presented here, spherical symmetry is imposed.
Consequently nuclear states can be characterized by their
angular momentum J and their parity $\pi$.
The individual Hartree-Fock wave functions are expanded on finite
sets of spherical harmonic oscillator (HO) wave-functions
containing 15 major shells for all nuclei.
For each nucleus, the value of the parameter $\hbar\omega$ of the HO
basis is taken as the one minimizing the HF total nuclear energy.

The RPA equations  are solved in matrix form in the p-h
representation. RPA energies do not appear very sensitive to the value
adopted for the HO parameter of the basis. For instance, by changing
the optimal HF value $\hbar\omega=8.7$ MeV in $^{208}$Pb by $10\%$, the
variation of the ISGMR energy ($13.46$ MeV) is less than $.5\%$ and
the energy of the first $2^+$ at $4.609$ MeV is changed by less than
$5$ keV.

Electric transition operators are defined according to:
\begin{equation}
\widehat{Q}_{JM} =\dspt\frac{e}{2}  \sum_{i}^{A} \left( 1 -\tau_z( i ) \right)
j_J(q r_i) \, Y_{JM} (\theta_i,\phi_i),
\end{equation}
where $j_J$ is a spherical Bessel function of order $J$, $q$ a
transferred momentum, $\tau_z$ the third component of the nucleon
isospin and $Y_{JM}$ the usual spherical harmonics.

The degree of collectivity of the excited states is measured from their
contribution to  the Energy Weighted Sum Rule (EWSR)
\begin{equation}
M_1( \widehat{Q}_{JM} ) = \sum_N ( E_N -E_0 ) \vert \langle N \vert \widehat{Q}_{JM}\vert
0 \rangle \vert ^2
\label{e4}\end{equation}
where $\vert 0 \rangle$ and $\vert N \rangle$ are the RPA correlated
ground state and excited states, respectively and $E_N -E_0$ their
excitation energies.
Eq.(\ref{e4}) can also be expressed as the average in the HF ground
state $\vert HF \rangle$ of a double commutator~\cite{Lipparini}:
\begin{equation}
\dspt M_1( \widehat{Q}_{JM} )=\frac{1}{2}\langle HF \vert \left[
\widehat{Q}_{JM}, \left[ \widehat{H},
\widehat{Q}_{JM} \right] \right]\vert HF \rangle.
\label{e3}\end{equation}

Therefore, exact values of $M_1( \widehat{Q}_{JM} )$ can be computed from
expression (\ref{e3}) whereas smaller values will be obtained from
(\ref{e4}), reflecting the finiteness of the particle-hole space used in the
RPA calculations.

\begin{figure}[!ht]
\centerline{\hbox{\psfig{file=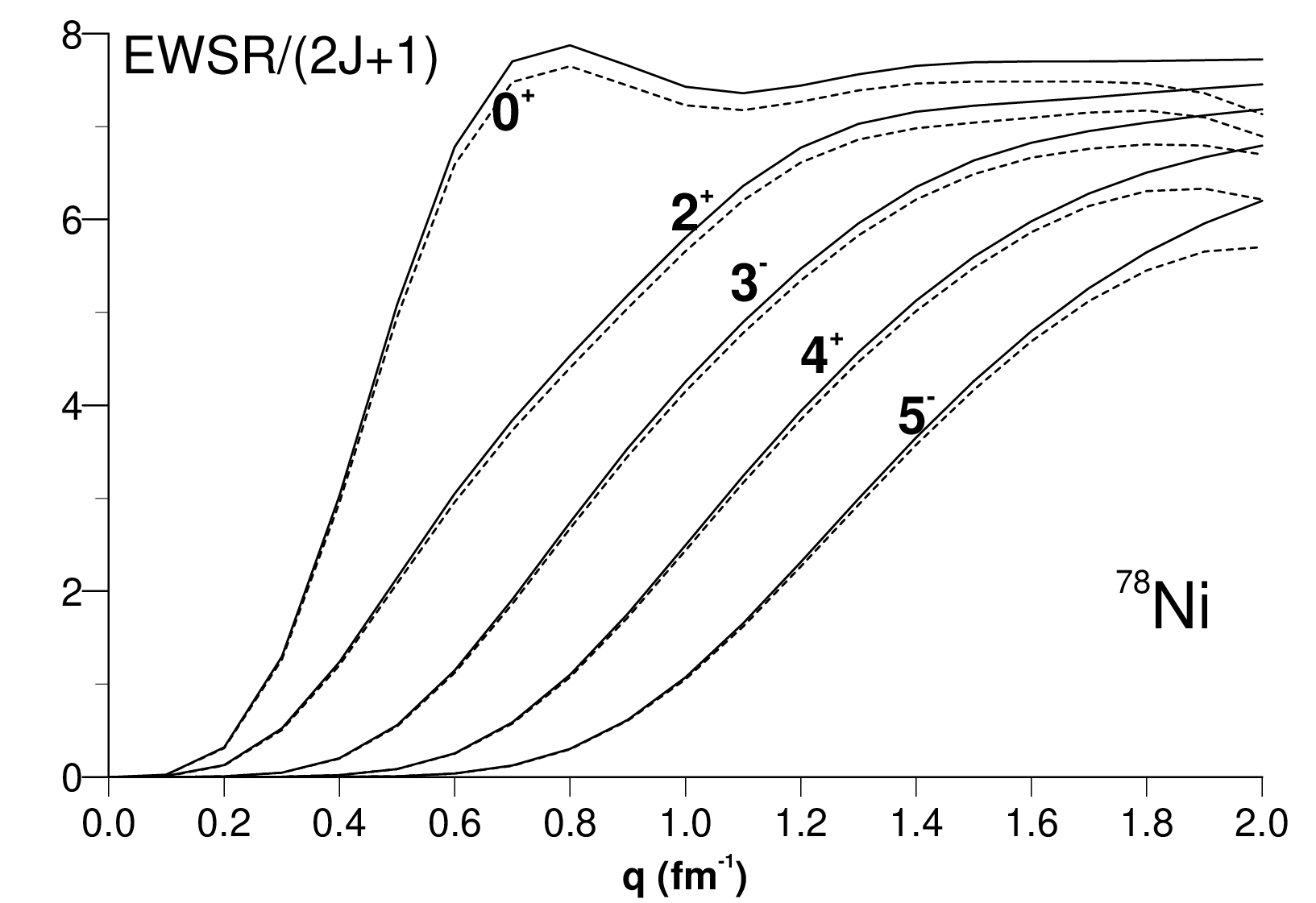,height=7.cm,angle=270}}}
\caption{Comparisons between the exact EWSR of eq.(\ref{e3}) (solid
line) and those deduced from eq.(\ref{e4}) (dotted line) in $^{78}$Ni
for the RPA states with  $J^\pi=$ $0^+$, $2^+$, $3^-$, $4^+$ and $5^-$. The
unit of the EWSR scale is e$^2$ MeV. The abscissa q represents the transferred
momentum}
\label{EWSR78Ni}
\end{figure}

A comparison between the values calculated from (\ref{e4}) and
(\ref{e3}) is shown in Figure \ref{EWSR78Ni} for $^{78}$Ni as an
example.
As can be seen, with the 15 major shell basis employed, RPA
calculations are able to describe with a reasonable accuracy the
nuclear response for $J^\pi=0^+$, $2^+$, $3^-$, $4^+$ and $5^-$ up to
transferred momenta $q$=1.5 fm$^{-1}$.

 \section{Results}
\begin{figure}
\centerline{\hbox{\psfig{file=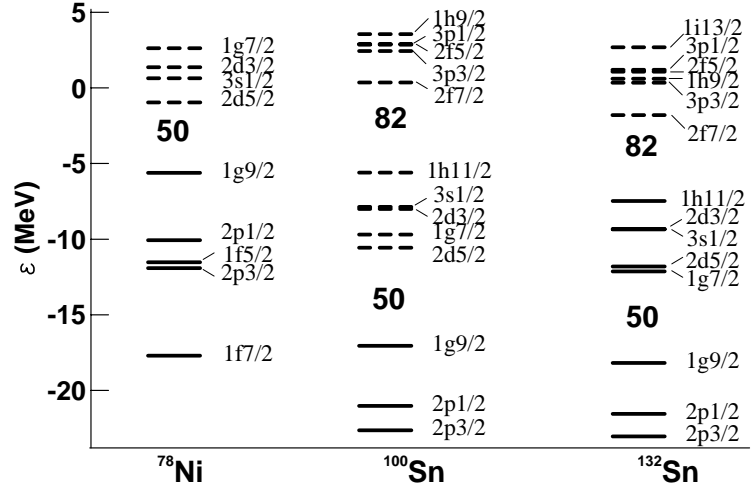,height=7.cm,angle=270}}}
\caption{ Single particle levels in the vicinity of the Fermi
surface for neutrons in the three studied exotic nuclei.
Filled and empty levels are represented by full and dashed lines, respectively.
The labels indicate the quantum numbers $(nlj)$ of the levels.}
\label{nivneutron}
\end{figure}

  First, we will discuss the validity of the doubly-magic nature of
these exotic nuclei.  The single-particle neutron spectra obtained in
$^{78}$Ni, $^{100}$Sn and $^{132}$Sn are shown in
Figure~\ref{nivneutron}. The N=50 gap in $^{78}$Ni and $^{100}$Sn and the
N=82 one in $^{132}$Sn are of the order of 5 MeV, which is less than
20$\%$ smaller than the gaps obtained for stable spherical nuclei with
same neutron numbers.
The same is true for the proton gaps at Z=28 in $^{78}$Ni and at Z=50
in tin isotopes.
That is, no significant reduction of the magic gaps are observed in
these  nuclei.  Therefore, the three exotic nuclei are still doubly
magic ones and the HF+RPA method is applicable to them.

In what follows, results for states with multipolarities $0^+$, $2^+$,
$1^-$ and $3^-$ are presented for four nuclei  $^{78}$Ni,
$^{100}$Sn  $^{132}$Sn and $^{208}$Pb, the latter nucleus being
included as a reference.

The strengths shown in the Figures are given in percentage of the EWSR
calculated in the long wavelength limit $q \rightarrow 0$.
The relevant formulas to be used in this limit for the different
values of $J$ are given in the appendix of Ref.~\cite{ref2}.

In the present calculations the continuum spectrum of the HF
Hamiltonian is approximated by a discrete one. As a consequence, the
RPA strength functions appear in the form of discrete peaks.
In order to make comparisons with experiments more meaningful,
energy centroids will be defined in terms of the moments
\begin{equation}
M_k\left( \widehat{Q}_{JM} \right) = \sum_N ( E_N -E_0 )^k \vert
\langle N \vert \widehat{Q}_{JM}\vert 0 \rangle \vert ^2 .
\label{eb4}\end{equation}
of the strength function. Two of these centroids will be used in
the following:  the mean value of the energy $M_1/
M_0$, and the so-called ``hydrodynamic" energy $\sqrt{M_1/ M_{-1}}$ for
isoscalar monopole resonances.

As experimental data on GR energies is scarce in exotic nuclei,
comparisons will often be made with the systematic $A^{-1/3}$
empirical laws approximately verified in stable nuclei~\cite{book2}.
Values from these systematics as well as available experimental data
are given in the Tables.

\subsection{Monopole states}

\begin{figure}
\centerline{\hbox{\psfig{file=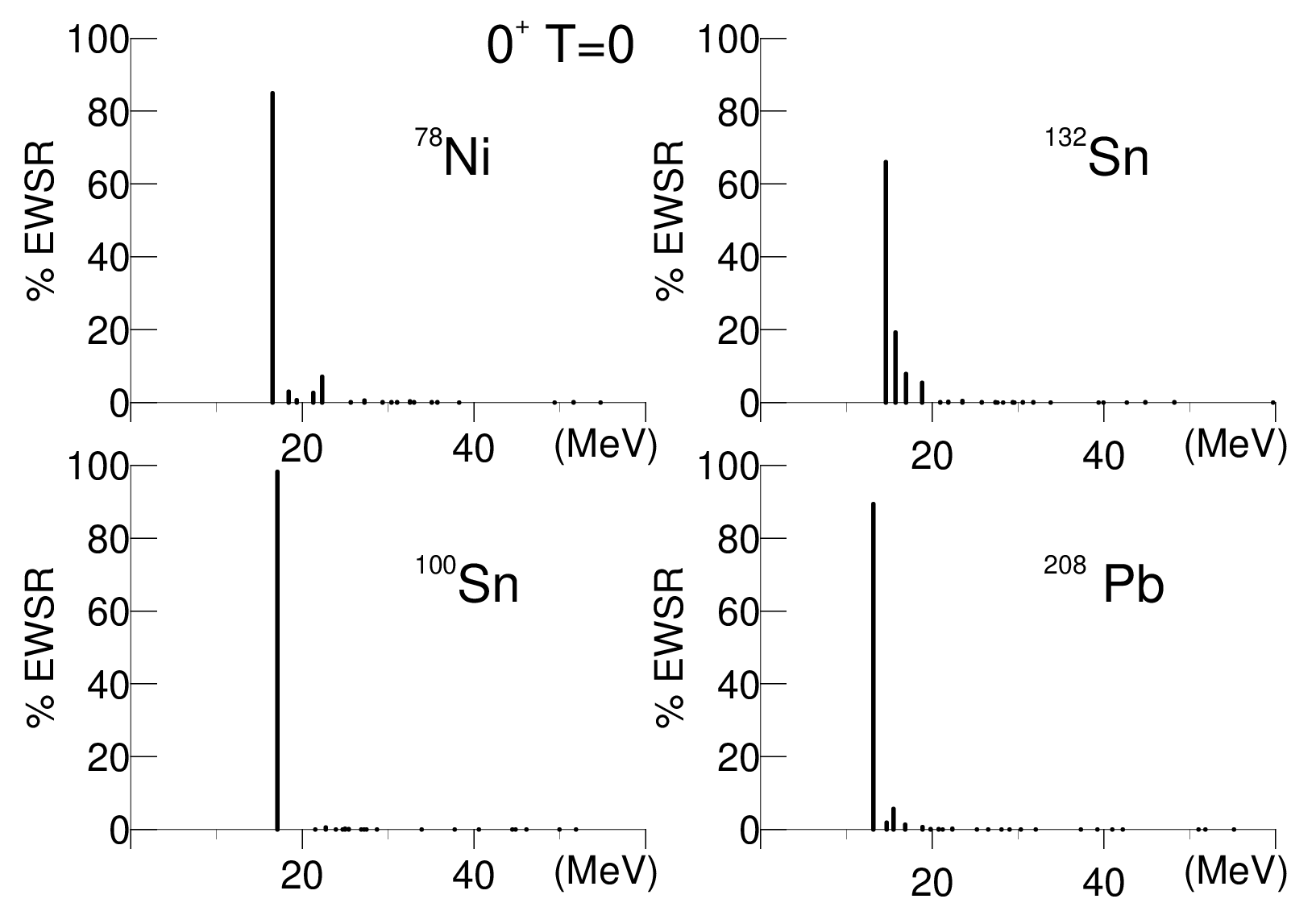,height=8.cm,angle=270}}}
\caption{Fraction of the EWSR carried by isoscalar $J^\pi=0^+$ states
in the four studied nuclei.}
\label{J0}
\end{figure}

 Figure~\ref{J0} and Table~\ref{tabd1sp} display the results obtained for
the Isoscalar Giant Monopole Resonance (ISGMR).

As is well known, the excitation energies of this resonance strongly
depends on the compression modulus $K_{nm}$ calculated in infinite
nuclear matter~\cite{blaizot}.
One observes in Table~\ref{tabd1sp} that the theoretical energies in
$^{208}$Pb, although in good agreement with the empirical $ 80
A^{-1/3}$ law, are 5\% lower than the experimental value of
Ref.~\cite{youn}. This difference is consistent with the compression
modulus found  in infinite nuclear matter with D1S',
 $K_{nm}$=209 MeV, which is slightly outside the interval 220-235 MeV
that explains the bulk of experimental data within non-relativistic
approaches~\cite{colo}.

Concerning  the three exotic nuclei, we note that resonance energies
significantly differ from the empirical law only in $^{78}$Ni.
It must be
noted that, of all three nuclei, $^{78}$Ni is the one where the squared neutron-proton
asymetry $\left(\left(N-Z\right)/A\right)^2$  most differs from the one of the stable
isotope: $\left(\left(N-Z\right)/A\right)^2-\left(\left(N-Z\right)/A\right)_{stable}^2$=0.78, 0.36
and -0.23 in $^{78}$Ni, $^{132}$Sn and $^{100}$Sn, respectively.
It is therefore tempting
to correlate the $\simeq$ 1.5 MeV lowering of the ISGMR found in $^{78}$Ni
with this large neutron excess, the contribution of the symmetry
term $K_{sym}$ to the finite nucleus incompressibility $K_A$ being negative
\cite{colo1,hiro}.

The strengths displayed in Figure~\ref{J0} show that the major part of the EWSR
is concentrated in a single peak in all four nuclei. This feature explains
why the two sets of theoretical energies listed in Table~\ref{tabd1sp} are
very close to each other. One notes that the fragmentation of the strength is
almost zero in the $N$=$Z$ nucleus $^{100}$Sn, whereas it is slightly bigger
in the other three nuclei which have neutron-proton asymmetry $(N-Z)/A$ in
the range .21--.28.

In Table~\ref{tab2}, we show the values of the mean monopole energies
$M_1/M_0$ obtained when different terms of the residual particle-hole (p-h)
interaction are left out of the RPA calculation. Columns $(1)$, $(2)$ and
$(3)$ refer to the mean energies calculated by leaving out the spin-orbit and
the Coulomb terms, the Coulomb term and the spin-orbit term, respectively.

One observes that the spin-orbit part of the residual interaction gives
a contribution to ISGMR energies ranging from 8\% in $^{78}$Ni to 5\%
in $^{208}$Pb. In contrast, the Coulomb contribution is larger in Pb
(3\%) and almost negligible in Ni.
These results are consistent with those discussed in Ref. \cite{colo} where
$^{40}$Ca, $^{90}$Zr and $^{208}$Pb were analyzed with the SLy4 interaction.
In the latter work, the inclusion in the constrained HF (CHF)
of the Coulomb force and of the spin-orbit
component of the Skyrme interaction was proved to be essential in order to
reconcile the value of $K_{nm}$ obtained with the Skyrme and Gogny forces.

\subsection{Quadrupole states}

\begin{figure}
\centerline{\hbox{\psfig{file=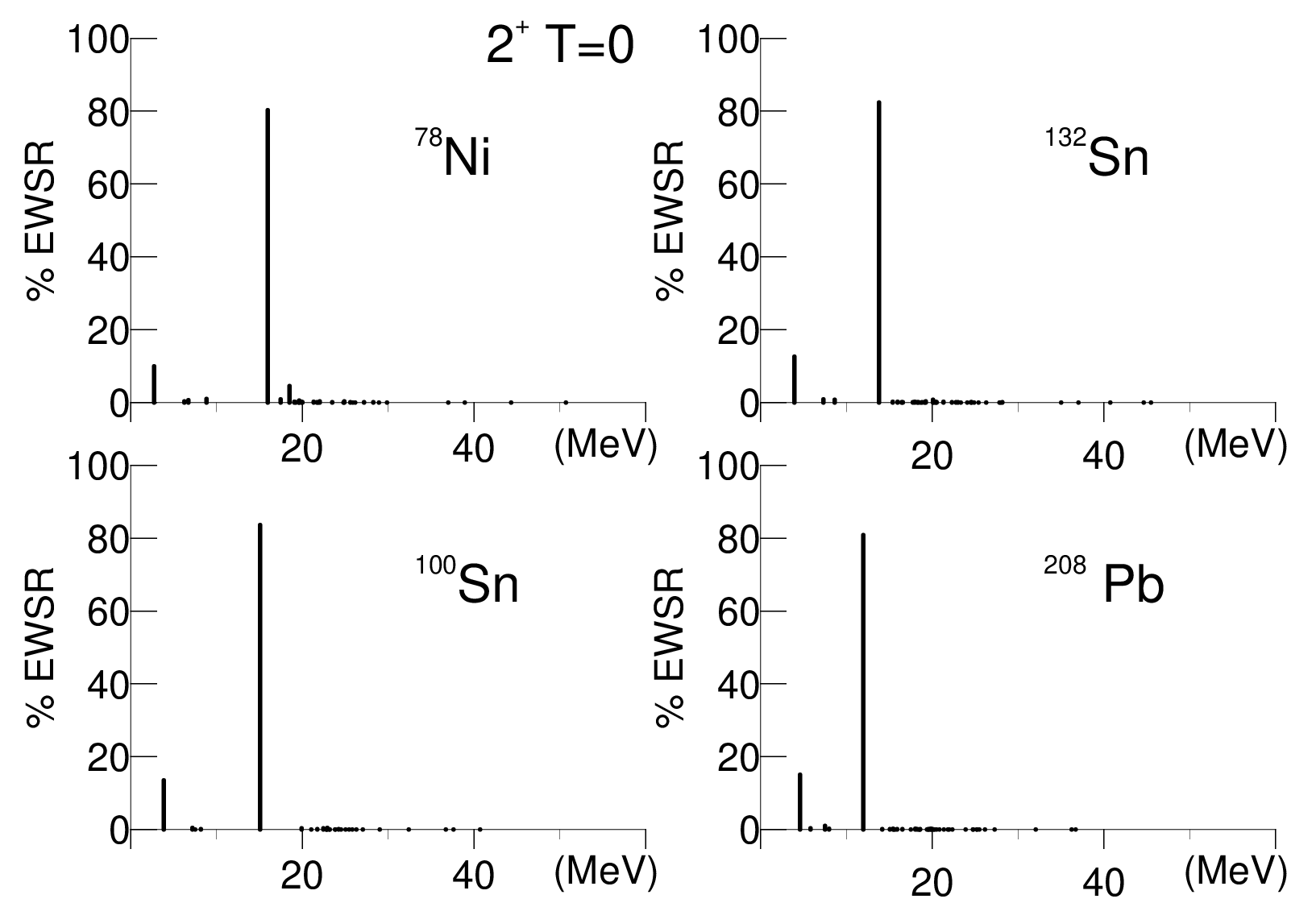,height=8.cm,angle=270}}}
\caption{Fraction of the EWSR carried by isoscalar $J^\pi=2^+$ states
in the four studied nuclei.}
\label{J2}
\end{figure}

Figure~\ref{J2} and Tables~ \ref{tab2D1Sp}, \ref{tab2D1Spp} and \ref{tab2p}
display the results obtained for isoscalar quadrupole states.
Figure~\ref{J2} shows that in all four nuclei the quadrupole strength
is divided essentially between two states: the isoscalar Giant
Quadrupole Resonance (ISGQR) exhausting $\simeq$~80\% of the EWSR with an
energy in the range 12-16 MeV and a lower-lying state at $\simeq$~3-5 MeV
carrying $\simeq$~10\%-15\% of the quadrupole strength. We will label the latter
$2^+_1$.

The theoretical ISGQR energies are calculated using $M_1/M_0$ excluding the $2^+_1$ state.
The results shown in Table~\ref{tab2D1Sp} are seen to be higher than
the  $A^{-1/3}$ systematics by 1.0--1.5 MeV. As the latter agrees well
with the experimental value in $^{208}$Pb, it is difficult to draw
definite conclusions concerning the behaviour of our results in the
three exotic nuclei. Let us mention that such large ISGQR energies can
be understood from a too large spreading of the particle-hole spectrum
in the $2^+$ channel at high energies. Such spreading is a consequence
of the value of the effective mass of the D1S' interaction (m$^*$/m =
0.7) which is the one giving correct single-particle properties in
mean-field calculations.  As is well known, taking into account the
coupling of RPA configurations to 2-particle--2-hole (2p-2h) states
would reduce this disagreement \cite{npa371,rmp}.
Clearly, such a coupling should be introduced in the present
calculations before reliable predictions for the ISGQR in exotic nuclei
can be made~\cite{Ghi}.  Let us mention that  the same is
true for the other giant resonances, with some dependence on the mode
quantum numbers~\cite{rmp}.
Nevertheless, few results have been obtained up to now with such
a coupling and it is difficult to
foresee the magnitude of energy shifts,
except for quadrupole and dipole states.

Our theoretical results for low-lying $2^+_1$ states  are
presented in Table~\ref{tab2D1Spp}. For these states, experimental data
exist both for $^{208}$Pb~\cite{Zie} and $^{132}$Sn~\cite{OR}.
As can be seen, a fair agreement between experiment and theory is
found  in $^{208}$Pb and an even better one in $^{132}$Sn, with B(E2)
values being of the same order of magnitude as experimental ones.
Let us point out that QRPA calculations applied to quadrupole states
have been made recently with the D1S interaction for a series of tin
isotopes including $^{132}$Sn~\cite{milan}.
In these calculations, the spin-orbit part and the coulomb part of the
residual interaction were omitted for simplicity reasons.
The $2^+$ energies were found larger than the
experimental ones by 400 keV in $^{102}$Sn and 1 Mev in $^{132}$Sn. The
 corresponding theoretical B(E2) values were lower than experimental ones
by at least a factor of two.

These results are consistent with those shown in
Table~\ref{tab2p} where the same quantities as those of Table~\ref{tab2D1Spp}
are displayed. They
have been calculated by leaving out from the D1S' p-h interaction the
spin-orbit and the Coulomb terms, the Coulomb term, the spin-orbit term and
no term, respectively. One observes that, as previously for monopole
vibrations, taking into account the spin-orbit part of the residual
interaction is essential to get results consistent with experimental data.

Going back to Table~\ref{tab2D1Spp}, $2^+_1$ energies are
similar in $^{100}$Sn and $^{132}$Sn, whereas a comparatively low value
is predicted in $^{78}$Ni.
 Let us note that the $2^+_1$ state in $^{78}$Ni is still higher
than the one in $^{56}$Ni, the other doubly magic Ni isotope, where the
experimental value of the $2_1^+$state is  2.7~MeV and the RPA
calculated one is 2.42~MeV with D1S'.

\begin{figure}
\centerline{\hbox{\psfig{file=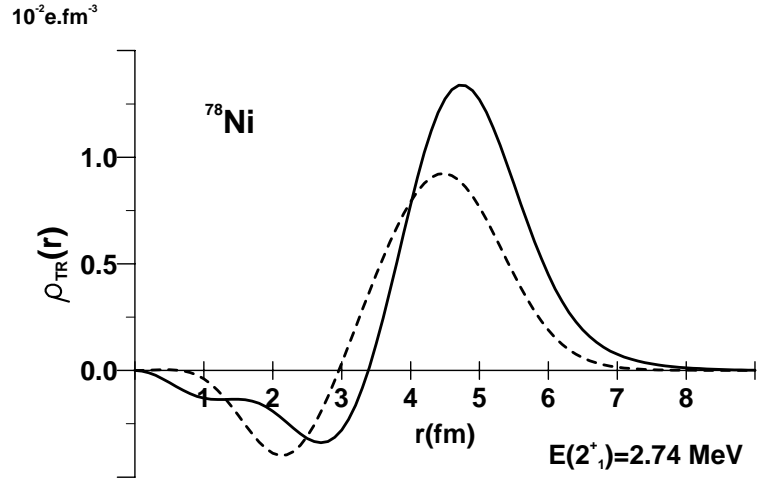,height=7.cm,angle=270}}}
\caption{Neutron (full line) and proton (dashed line) transition
densities for the first $2^+$ state in $^{78}$Ni.}
\label{denstr}
\end{figure}

The collectivity of this $2^+$ state appears larger in
$^{100}$Sn than in $^{132}$Sn and rather weak in $^{78}$Ni.
Figure~\ref{denstr} displays the transition density $\rho_{TR}$ of
this first $2^+_1$ state in $^{78}$Ni.
The definition of the transition density is the same as the one given in appendix of
Ref.~\cite{ref2}.
One observes that the  two transition densities are in phase
 and that the neutron transition density is higher than the proton one
and displaced to a larger radius.
This mode can therefore be interpreted as an isoscalar surface mode
dominated by neutron excitation.

\subsection{Dipole states}

\begin{figure}
\centerline{\hbox{\psfig{file=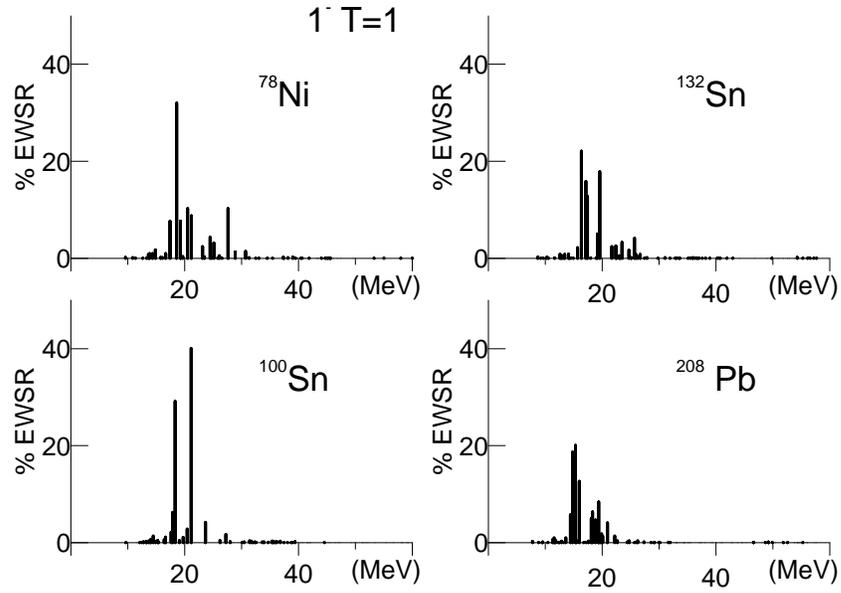,height=8.cm,angle=270}}}
\caption{Fraction of the EWSR carried by isovector $J^\pi=1^-$ states
in the four studied nuclei.}
\label{J1}
\end{figure}

Results for the isovector dipole resonance (IVGDR) are presented in
Figure~\ref{J1} and Table~\ref{tabDip}.
$^{100}$Sn is the nucleus where the giant dipole mode is the least
fragmented with 70\% of the strength concentrated into two peaks.
The dipole responses of $^{208}$Pb and $^{132}$Sn and to a lesser
extent of $^{78}$Ni also appear concentrated into two main energy
regions. It is expected, that the fragmentation is somewhat reduced by
the coupling of the RPA modes to 2p--2h states, producing smoother
strength functions, as in Refs.~\cite{colo2,colo3} where Skyrme forces
were used.

In $^{100}$Sn the mean value $M_1 / M_0 =$ 19.98 MeV is  3
MeV larger than the systematic 79$A^{-1/3}$ law (17.02 MeV).
The EWSR value given in Thomas-Reiche-Kuhn (TRK) unit is 1.59, which
is large compared to typical experimental values~\cite{refexp}.
The IVGDR in $^{132}$Sn is more fragmented than in $^{100}$Sn. As in
$^{100}$Sn the mean energy value, 18.33 MeV, is much larger than
systematics  (79$A^{-1/3}=$ 15.52 MeV) and the EWSR value is 1.58.
In the case of $^{78}$Ni, the IVGDR is quite fragmented with one major
peak and  smaller ones at higher energy. The mean energy value, 20.31
MeV, remains higher than systematics  (79$A^{-1/3}=$ 18.49 MeV) and the
EWSR in TRK unit is 1.57.

It must be said that IVGDR excitation energies calculated with the
Gogny force usually overestimate experimental data. In the case of
$^{208}$Pb, the calculated mean value is 16.50 MeV, which is quite large
compared to experiment (13.43 MeV~\cite{refexp}), but smaller than the result
of Ref.~\cite{ref2}.
Let us note that, ignoring the higher part of the IVGDR response by
keeping only the strength around the main lower energy peak,
considerably improves the agreement with systematic estimations : mean
energy values become 19.28 MeV, 18.16 MeV, 16.81 MeV and 14.99 MeV in
$^{78}$Ni, $^{100}$Sn, $^{132}$Sn and  $^{208}$Pb, respectively.

 In fact, calculated IVGDR energies and EWSR appear quite sensitive
to the energy interval considered and also to the components of the
effective interaction included in the p-h residual interaction.
This is shown in Table~\ref{newtable} where mean IVGDR energies and
EWSR in $^{208}$Pb are listed for three energy integration intervals
and for RPA calculations where Coulomb and/or spin-orbit terms are not
included in the RPA matrix elements.
One can see that the overestimation obtained with the Gogny force
decreases by $\simeq$ 700 keV when the Coulomb and the spin-orbit forces
are ignored, which is usually done in RPA calculations employing
Skyrme forces, see however Ref.~\cite{Jun}.
By taking all the terms of the Gogny force and considering the
largest energy interval, the calculated EWSR given is 1.59 in TRK units.
This value is higher than the experimental one obtained for
a 10-20 MeV energy interval (1.37)~\cite{refexp} but lower than the one
obtained for a energy interval going up to 140 MeV
(1.78)~\cite{refexp26}. In this case, however, another mechanism,
the "quasideuteron effect", is expected to play a major role in the photon
absorption~\cite{refexp26}.

It is of great interest, beyond nuclear physics itself, to study the
amount of excited low-lying dipole strength, that is the often
called "pygmy" resonances.
In terms of EWSR, we obtain much less than 1\% strength below 10
MeV in Ni and Sn nuclei, and about that amount in $^{208}$Pb.
The result for Pb is in agreement with the data of Ref.~\cite{Rich}.
The absence of collective states in the low-lying region is at variance
with the results of relativistic RPA calculations~\cite{Vre}, but
agrees with the arguing in Ref.~\cite{hiro1}. There, it is pointed out
that the soft dipole strength should decrease in  nuclei displaying
a neutron skin, compared to that in light halo nuclei because of a
more efficient coupling to the IVGDR. On the other hand, the coupling
to 2p--2h can significantly increase the amount of low-lying
strength~\cite{colo2,colo3}.

 By introducing a very small renormalization factor (1.01-1.03) of
the residual interaction the isoscalar spurious mode can be made to
appear at zero frequency.
This factor is introduced only in the  $J^{\pi}=1^-$ subspace.
In Table~\ref{tab1}, the values of the energy of this state are shown
as calculated with or without different parts of the D1S' p-h
interaction. For each nucleus the same renormalisation factor
is used in the four cases.
The symbol $\in \Im$ means that the RPA eigenvalue is imaginary.  These
results show, as expected, that the consistency between the HF field
and the residual interaction is important for the treatment of the
spurious states.

\subsection{Octupole states}

\begin{figure}
\centerline{\hbox{\psfig{file=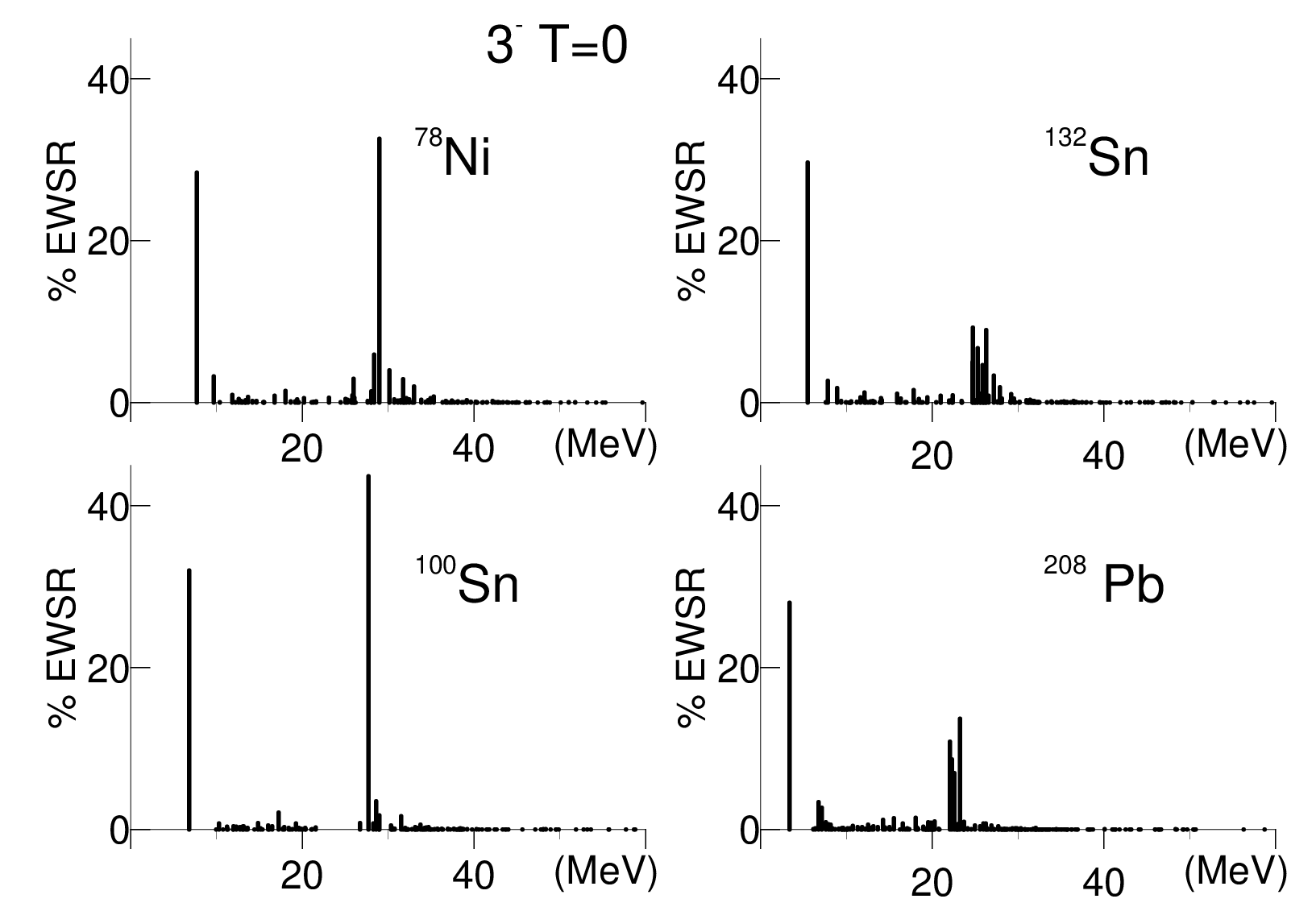,height=8.cm,angle=270}}}
\caption{Fraction of the EWSR carried by isoscalar $J^\pi=3^-$ states
in the four studied nuclei.}
\label{J3}
\end{figure}

As shown in Figure~\ref{J3}, the $J^\pi=3^-$ states belong to two
well-separated energy regions. Only the component at energies larger
than $\simeq$15 MeV can be considered as a genuine giant resonance, the High
Energy Octupole Resonance (HEOR).
Keeping only high energy regions (19-35 MeV  for $^{100}$Sn, 22-31 MeV
for $^{132}$Sn, 22-44 MeV for $^{78}$Ni and 13-28 MeV for $^{208}$Pb),
the mean calculated HEOR energies are  28.16 MeV, 26.06 MeV, 29.51 MeV
and 23.20 MeV, respectively.
These values give systematics $E_0 A^{-1/3}$, with $E_0 =$ 130, 132, 126, and
137 in the four nuclei, to be compared with the usual estimate
$110 A^{-1/3}$~\cite{refexp2}.
Previous studies in stable nuclei~\cite{ref2} gave values between
$130 A^{-1/3}$ and $140 A^{-1/3}$ for heavy nuclei and around
$120 A^{-1/3}$ in lighter ones.
We therefore do not observe a strongly different behaviour of HEOR
energies in exotic nuclei compared to the one previously obtained
along the valley of stability.

The characteristics of the low energy $3^-$ states are reported in
Table~\ref{tab3D1Sp}. The influence of the different components of
the D1S' force included in the p-h interaction is also shown.
The effect of the spin-orbit term appears to be smaller than for the
quadrupole states in Table 5, especially for $^{78}$Ni.

\subsection{Isovector strength}

\begin{figure}
\centerline{\hbox{\psfig{file=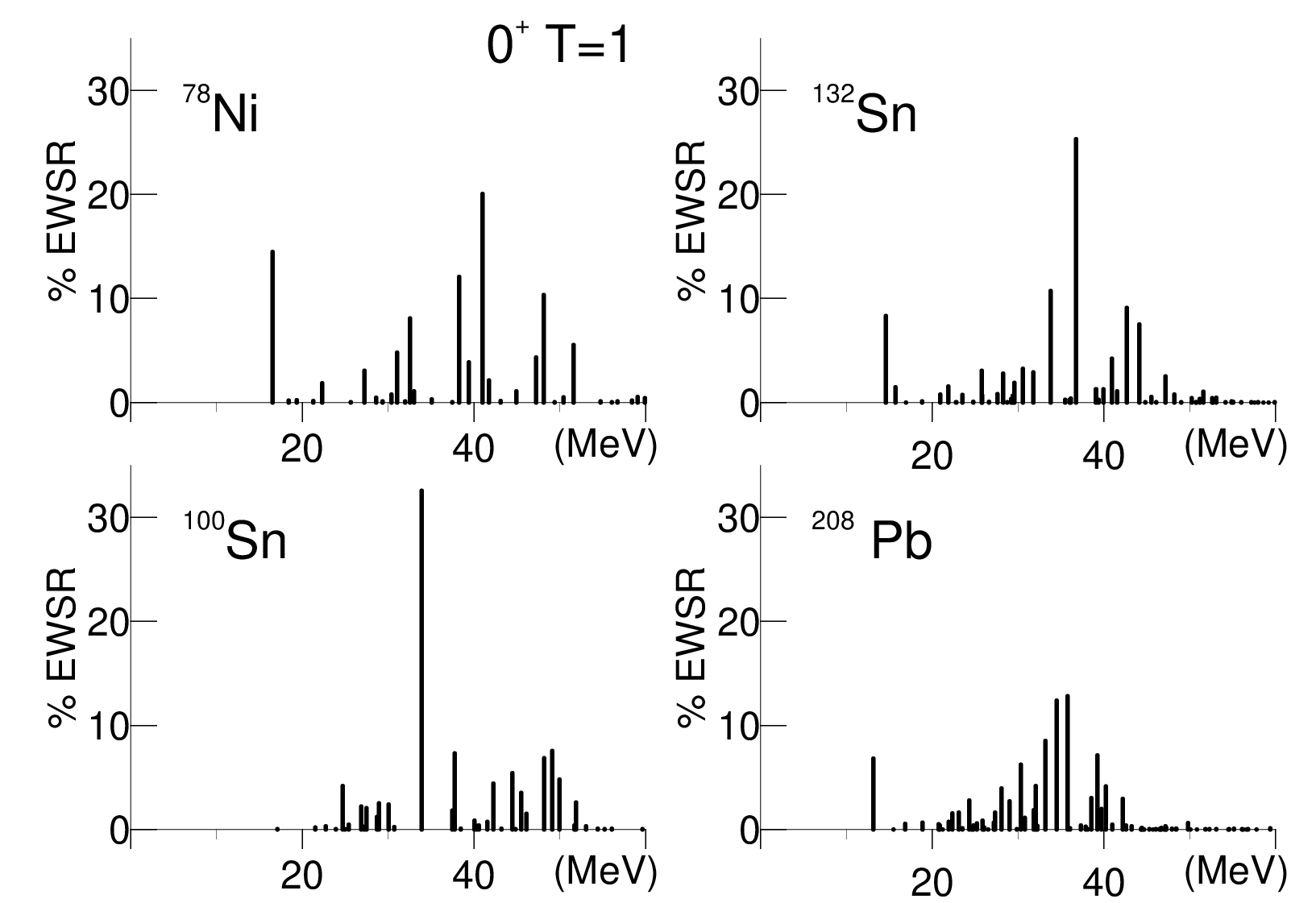,height=8.cm,angle=270}}}
\caption{Fraction of the isovector EWSR carried by $J^\pi=0^+$ states
in the four nuclei.}
\label{JV0}
\end{figure}

\begin{figure}
\centerline{\hbox{\psfig{file=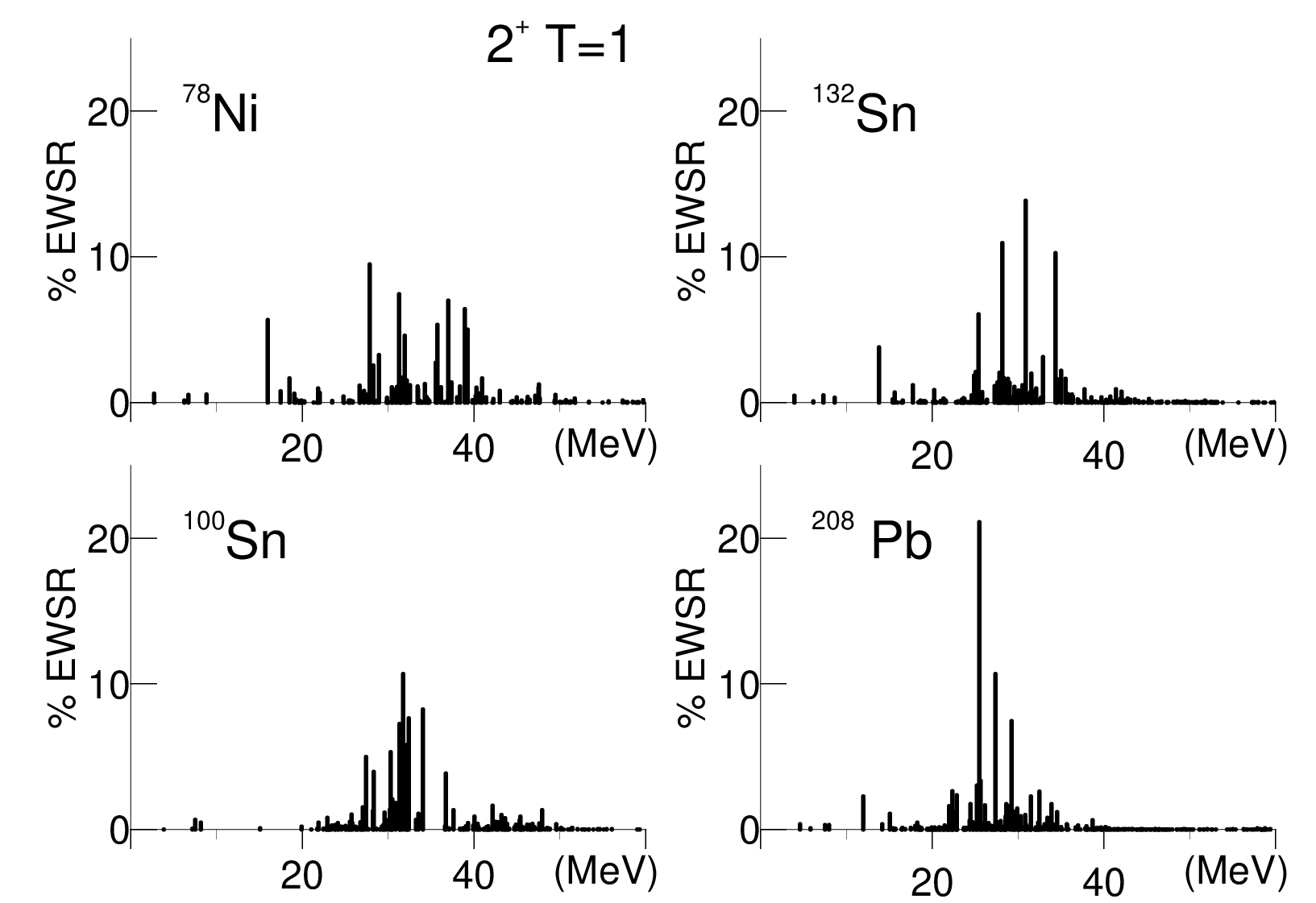,height=8.cm,angle=270}}}
\caption{Fraction of the isovector EWSR carried by $J^\pi=2^+$ states
in the four nuclei.}
\label{JV2}
\end{figure}

\begin{figure}
\centerline{\hbox{\psfig{file=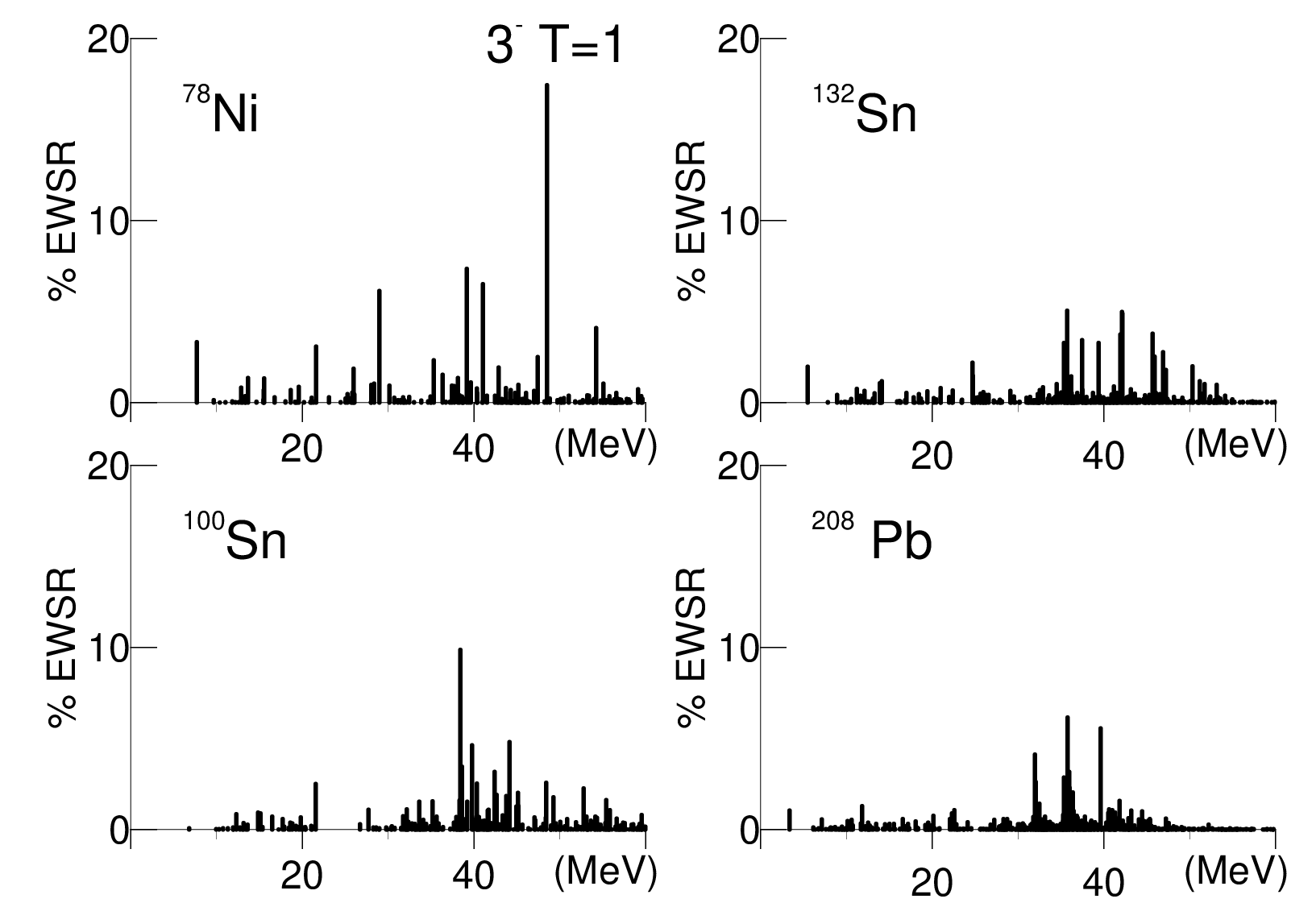,height=8.cm,angle=270}}}
\caption{Fraction of the isovector EWSR carried by $J^\pi=3^-$ states
in the four nuclei.}
\label{JV3}
\end{figure}

In Figures~\ref{JV0}--\ref{JV3}, the fractions of the isovector EWSR
carried by the $J^\pi=$ $0^+$, $2^+$, $3^-$ states is drawn.
In this case, systematics for stable nuclei are not yet well
known~\cite{book2} and is not reported.
Note that only the transition operator is
changed compared to the isoscalar case in Figures \ref{J0}, \ref{J2} and \ref{J3}.
From the comparison between the two sets of figures, a much larger
fragmentation of the strength is found in the isovector case, and a
mixed (isoscalar-isovector) character of several states appears, as
expected, in particular in $^{78}$Ni.

\section{Conclusion}

To summarize, we have presented the results obtained for different
giant resonances in three  doubly magic exotic nuclei, using the HF+RPA
approach and the Gogny force.
The largest difference with usual doubly magic nuclei inside the valley
of stability occurs in $^{78}$Ni where the ISGMR appears significantly
lower than systematics. This seems to be due to the large
proton-neutron asymmetry of this nucleus.

The fragmentation of the isovector dipole strength has to be
explored further in order to see the correlation or the no-correlation
with proton-neutron radius differences.
 In particular, the nature of the double-peaks obtained in tin
isotopes remains to be determined.

Results obtained in the three exotic nuclei for the ISGQR and HEOR
resonances are similar to  those of $^{208}$Pb, but more exotic
systems have to be studied to confirm such a trend.

Low energy states and B(E2) values appear to be well reproduced
 within the present approach, in particular the first  $2^+$ in
$^{132}$Sn.

 From a more general point of view, we have found that the
spin-orbit component of the p-h residual interaction plays a very
important role in the structure of the low-lying quadrupole and
octupole states, as it strongly influences both excitation energies and
transition probabilities.
 Similarly, our results show that including the Coulomb force in the
RPA p-h matrix elements significantly affects
IVGDR energies and EWSR.

\section{Acknowledgments}

The authors want to thank D. Gogny for his interest in this work and
useful comments.
P.F.B. acknowledges  the Service de Physique Nucl\'eaire,
CEA/DAM--Ile--de--France at Bruy\`eres--le--Ch\^atel for financial
support and warm hospitality during the periods in which parts of this
work were performed.

\pagebreak


\vspace{2cm}

\pagebreak
\begin{table}
\begin{center}
\begin{tabular}{|c|c|c|c|c|}
\hline
$0^+$ $T$=0&$\dspt \frac{M_1}{M_0}$ & $\dspt\sqrt{\frac{M_1}{M_{-1}}}$
&80 A$^{-1/3}$                      & Exp    \\
\hline
$^{78}$Ni  &  17.17&  17.07&   18.72  &       \\
$^{100}$Sn &  17.22&  17.18&   17.23  &       \\
$^{132}$Sn &  15.29&  15.22&   15.72  &       \\
$^{208}$Pb &  13.46&  13.42&   13.50  & 14.17$\pm$ 0.28  \\
\hline
\end{tabular}
\caption{ Mean values and ``hydrodynamic'' centroids  of ISGMR energies
in MeV obtained with the D1S' force in the four studied nuclei compared with the
empirical $ 80 A^{-1/3}$ law and the $^{208}$Pb experimental value from
Ref.\cite{youn}.}
\label{tabd1sp}

\vspace{1cm}

\begin{tabular}{|c|c|c|c|c|}
\hline
$M_1/M_0$\ & (1)    &(2)     &(3)  &(tot)    \\
\hline
$^{78}$Ni  & 18.55   &17.10    & 18.59   & 17.17    \\
$^{100}$Sn & 18.19   &16.81    &18.54    & 17.22     \\
$^{132}$Sn & 16.07   &15.06    & 16.26   & 15.29    \\
$^{208}$Pb & 13.73   &13.05    & 14.10   & 13.46       \\
\hline
\end{tabular}
\caption{Mean ISGMR energies in MeV obtained by leaving out from the D1S' p-h
interaction:
(1) the spin-orbit and the Coulomb terms, (2) the Coulomb term,
(3) the spin-orbit term, (tot) no term.}
\label{tab2}
\end{center}
\end{table}

\begin{table}
\begin{center}
\begin{tabular}{|c|c|c|c|}
\hline
ISGQR &  D1S' & 64$A^{-1/3}$& Exp.  \\
\hline
$^{78}$Ni    & 15.94&  14.98 &  \\
$^{100}$Sn   & 15.13&  13.79 &  \\
$^{132}$Sn   & 13.79&  12.57 &  \\
$^{208}$Pb   & 11.98&  10.80 &10.60  \\
\hline
\end{tabular}
\caption{ Mean values of ISGQR energies in MeV obtained
with D1S' in the four studied nuclei compared with the
empirical $ 64 A^{-1/3}$ law and the $^{208}$Pb experimental value from
Ref. \cite{book2}.}
\label{tab2D1Sp}

\vspace{1cm}

\begin{tabular}{|c|cc|cc|}
\hline
 & & & Experiment& \\
$2^+_1$  &  E  &   B(E2) &  E(MeV)     &B(E2) \\
\hline
$^{78}$Ni   & 2.73 & 466 & &\\
$^{100}$Sn  & 3.84 & 1431& & \\
$^{132}$Sn  & 3.97 & 1134&  4.041& 1400 (600)\\
$^{208}$Pb  &4.609 & 2781& 4.08 &3180 (160) \\
\hline
\end{tabular}
\caption{ Energies in MeV and  corresponding B(E2) in e$^2$fm$^4$ of $2^+_1$ states calculated with the
D1S' interaction. Existing experimental data from
Refs. \cite{Zie} and \cite{OR} are also listed.}
\label{tab2D1Spp}

\vspace{1cm}

\begin{tabular}{|c|cc|cc|cc|cc|}
\hline
$2^+_1$   &(1)   &       &(2)    &       &(3)    &        &(tot) &          \\
          &E&B(E2)  &E &B(E2)  &E &B(E2)   &E& B(E2)     \\
\hline
$^{78}$Ni & 3.53  & 257  &2.84  &456  & 3.43  & 271   & 2.73 & 466        \\
$^{100}$Sn& 4.64  & 1103 &3.95  &1552 & 4.48  & 1041  & 3.84 & 1431      \\
$^{132}$Sn& 4.61  & 775  &4.04  &1182 & 4.53  & 770   & 3.97 & 1134      \\
$^{208}$Pb& 5.15  & 2305 &4.65  &3145 & 5.09  & 2123  & 4.61 & 2781       \\
\hline
\end{tabular}
\caption{ Energies in MeV and B(E2) of $2^+_1$ states obtained by leaving
out from the D1S' p-h interaction:
(1) the spin-orbit and the Coulomb terms, (2) the Coulomb term,
(3) the spin-orbit term, (tot) no term.}
\label{tab2p}
\end{center}
\end{table}

\begin{table}

\vspace{1cm}

\begin{center}
\begin{tabular}{|c|c|c|c|}
\hline
IVGDR &  D1S' & 79$A^{-1/3}$& Exp.  \\
\hline
$^{78}$Ni    &20.31  &18.49   &  \\
$^{100}$Sn   &19.98  &17.02   &  \\
$^{132}$Sn   &18.33  &15.52   &  \\
$^{208}$Pb   &16.50  &13.33   &13.43  \\
\hline
\end{tabular}
\caption{ Mean values of IVGDR energies in MeV obtained
with D1S' in the four studied nuclei compared with the
empirical $ 79 A^{-1/3}$ law and the $^{208}$Pb experimental value from
Ref. \cite{refexp}.}
\label{tabDip}

\vspace{1cm}

\begin{tabular}{|c|cc|cc|cc|cc|c|}
\hline
$^{208}Pb$             &(1)    &     &(2)    &      &(3)    &      &(tot)  &     & Exp.  \\
                       &$<E>$  &EWSR &$<E>$  &EWSR  &$<E>$  &EWSR  &$<E>$  & EWSR&EWSR   \\
\hline
$\left[ 0- 140\right]$ & 15.88 &1.63 & 15.70 & 1.62 & 16.71 & 1.59 & 16.50 & 1.59 & 1.78  \\
$\left[ 0- 20\right]$  & 15.10 &1.41 & 15.31 & 1.47 & 15.83 & 1.33 & 15.86 & 1.42 & \\
$\left[ 10- 20\right]$ & 15.20 &1.39 & 15.17 & 1.49 & 15.90 & 1.32 & 15.95 & 1.41 & 1.37 \\
\hline
\end{tabular}
\caption{ Mean IVGDR energies in MeV and EWSR in TRK units for
$^{208}$Pb calculated by leaving out from the D1S' p-h interaction:
(1) the spin-orbit and the Coulomb terms, (2) the Coulomb term,
(3) the spin-orbit term, (tot) no term.
The three lines show the results obtained for the three energy
intervals given in MeV in the leftmost columm.
The rightmost column gives experimental EWSR in TRK units.}
\label{newtable}

\vspace{1cm}

\begin{tabular}{|c|c|c|c|c|}
\hline
$1^-_{sp}$ T=0 & (1) & (2) & (3) & (tot) \\
\hline
$^{132}$Sn & $\in \Im$ & $\in \Im$ & 2205.78 & 4.26 \\
\hline
$^{208}$Pb & $\in \Im $ & $\in \Im$ & 1605.19 & 2.29   \\
\hline
\end{tabular}
\caption{Energy in keV of the isoscalar $1^-_{sp}$ spurious state calculated
by leaving out from the D1S' p-h interaction:
(1) the spin-orbit and the Coulomb terms, (2) the Coulomb term,
(3) the spin-orbit term, (tot) no term. The symbol $\in \Im$ means that the
RPA eigenvalue is imaginary.}
\label{tab1}

\vspace{1cm}

\begin{tabular}{|c|cc|cc|cc|cc|cc|}
\hline
$3^-_1$   &(1)   &       &(2)    &       &(3)    &        &(tot) &       &Exp &\\
          &E&B(E3)  &E &B(E3)  &E &B(E3)   &E& B(E3) &E&  B(E3)  \\
\hline
$^{78}$Ni & 7.95 &0.170 & 7.80  & 0.221& 7.87  & 0.181 & 7.70 & 0.231&     &     \\
$^{100}$Sn& 7.26 &0.130 & 6.95  & 0.149& 7.13  & 0.128 & 6.82 & 0.147&     &     \\
$^{132}$Sn& 5.78 &0.123 & 5.60  & 0.139& 5.72  & 0.124 & 5.53 & 0.140&     &      \\
$^{208}$Pb& 3.55 &0.725 & 3.38  & 0.782& 3.57  & 0.677 & 3.39 & 0.727& 2.6 & 0.611 (120)    \\
\hline
\end{tabular}
\caption{ Energies in MeV of the first $3^-$ state and corresponding B(E3) in
$10^{6}e^2 fm^6$ calculated by leaving out from the D1S' p-h
interaction:
(1) the spin-orbit and the Coulomb terms, (2) the Coulomb term,
(3) the spin-orbit term, (tot) no term.
Experimental data from Ref. \cite{spear} is also listed.}
\label{tab3D1Sp}
\end{center}
\end{table}

\end{document}